\documentclass[preprint,12pt,authoryear]{elsarticle}
\usepackage{amsmath}
\usepackage{amssymb}
\usepackage[export]{adjustbox}
\usepackage{verbatim}
\usepackage{xcolor}
\usepackage{wrapfig}
\usepackage{setspace}
\usepackage{pdfpages}
\usepackage{standalone}
\usepackage{todonotes}
\definecolor{shadecolor}{RGB}{240,255,240}
\usepackage{color}
\usepackage[inline]{enumitem}
\usepackage{graphicx}
\usepackage[space]{grffile}
\usepackage{latexsym}
\usepackage{amsfonts,amsmath,amssymb}
\usepackage{url}
\usepackage[T1]{fontenc}
\usepackage[utf8]{inputenc}
\usepackage{longtable}
\usepackage{multirow,booktabs}
\usepackage{sidecap}

\usepackage{hyperref}
\hypersetup{colorlinks=true,
					  allcolors = {blue},
}

\newcommand{\cotwo}{$\mathrm{CO}_2$}

\newcommand{\degree}{$^\circ$}

\journal{Icarus}

\begin{document}

\begin{frontmatter}


\title{How Martian araneiforms get their shapes: morphological analysis and diffusion-limited aggregation model for polar surface erosion}

\author[lasp]{Ganna Portyankina}
\author[psi]{Candice J. Hansen}
\author[lasp]{Klaus-Michael Aye}

\address[lasp]{Laboratory for Atmospheric and Space Physics, University of Colorado in Boulder, 3665 Discovery Drive, Boulder, CO, 80303}
\address[psi]{Planetary Science Institute, 1700 East Fort Lowell, Tucson, AZ 85719}

\begin{abstract}

Araneiforms are radially converging systems of branching troughs exhibiting fractal properties.
They are found exclusively in the Southern polar regions of Mars and believed to be result of multiple repetitions of cold \cotwo{} gas jets eruptions.
Araneiform troughs get carved by the overpressurized gas rushing underneath a seasonal ice layer towards a newly created opening.
Current work is  an attempt to quantitatively analyze araneiforms patterns and model their formation mechanism.
Araneiform shapes range from small and simple structures of one or two connected branches to intricate dendritic patterns several kilometers across.
The dendritic quality of most araneiforms are suggestive that they can be described in terms commonly applied to terrestrial rivers.
The main difference between rivers and araneiforms from the morphological point of view is that araneiform patterns are not directed or in any way influenced by the local gravitational gradient.
We have adapted and for the first time applied to Martian araneiforms qualitative morphological analysis typically used for terrestrial rivers. 
For several locations in the Martian Southern polar regions we have calculated the largest order of tributaries, tributary densities, and bifurcation ratios.
We have shown that the large and well-developed araneiforms (with tributary orders larger than 4) closely follow Horton’s law of tributary orders and have bifurcation ratio that falls well inside the range of terrestrial rivers.
We have implemented a two-dimensional Diffusion-Limited Aggregation (DLA) model that describes formation of dendrite shapes by mathematical probabilistic means. 
We compared modeled dendrite shapes to the araneiform shapes observed in the Martian polar regions and evaluated their similarity using the morphological analysis of araneiforms.
We showed that DLA model can successfully recreate 2D shapes of different observed araneiforms. 
To simulate the patterns that deviate from central symmetry it is required to modify the main governing parameter of DLA model: its two-dimensional probability field.
An analog to the governing probability field in the case of araneiforms is the gas pressure in the cavity underneath the ice layer right before the gas starts moving towards the newly created escape vent.
Thus modeling the creation of araneiform patterns with DLA leads to better understanding of seasonal processes that create them.

\end{abstract}


\end{frontmatter}



\section{Introduction: araneiforms and cold jet erosion}\label{Intro}

\begin{figure}[]
\includegraphics[width=0.95\columnwidth]{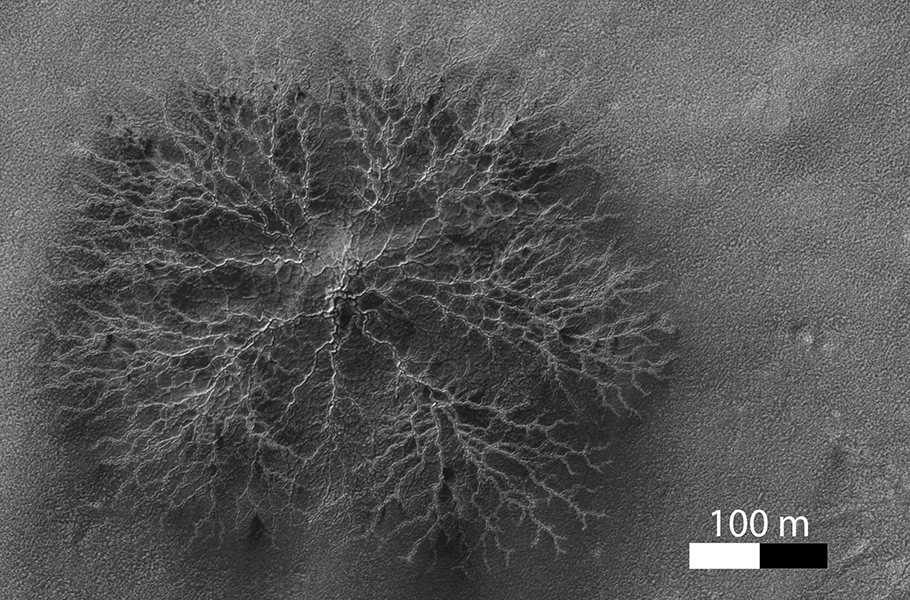}
\includegraphics[width=0.95\columnwidth]{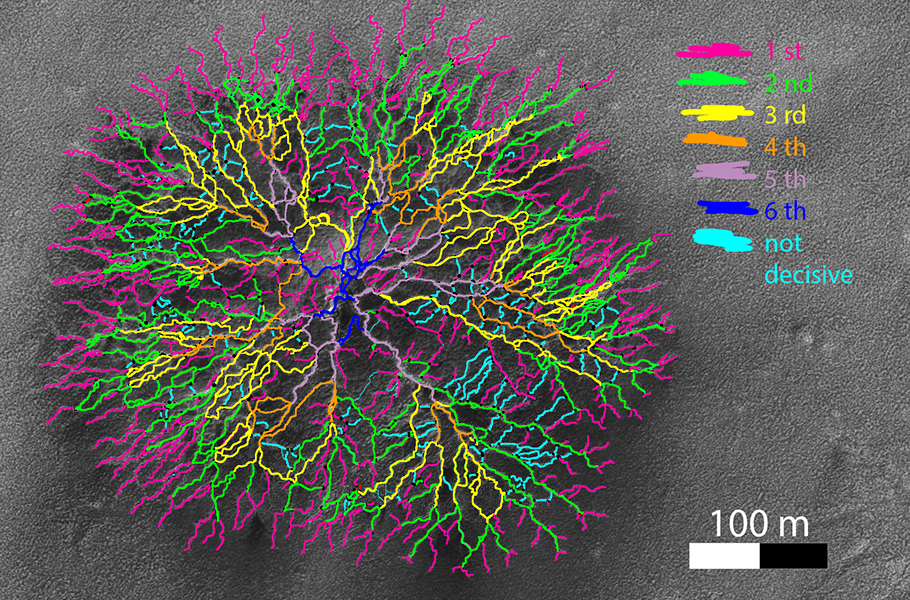}
\caption{\textbf{\label{fig:complex_spider}} Top: A subframe of HiRISE image \nolinkurl{ESP_047802_0980} showing a complex araneiform. Bottom: The same araneiform with its tributaries traced and colored accordighn to their order from 1 to 6. The ordering scheme used is Horton-Strahler stream order scheme with adaptations for Martian applications.
}
\end{figure}

According to the most accepted definition, araneiforms are radially converging systems of branching troughs exhibiting fractal properties.
Figure~\ref{fig:complex_spider} shows a complex araneiform imaged by High Resolution Imaging Science Experiment (HiRISE) on the Mars Reconnaissance Orbiter (MRO). 
The troughs of these systems are branching converging towards their centers similar to river tributaries converging downhill on their local topography.
Fig. \ref{fig:complex_spider} top panel is an image of a classical araneiform.
It is a well developed dendritic structure approximately 580~m across with multiple branching tributaries.
Bottom panel shows these tributaries traced and color-coded according to their orders calculated using the procedure described in section \ref{sec:dendrite_orders}.
Araneiforms are similar to rivers in morphology so that the methods used to study river pattern formation can be applied to araneiforms.
However, the caution must be taken in application of the same techniques because the formation mechanism for araneiforms is different to the on eof rivers: instead of water erosion araneiforms are created by erosive flux of \cotwo{} gas and have no relation to movement of any liquid under influence of gravity.

Araneiforms were first detected and colloquially named ''spiders'' by the Mars Orbiter Camera (MOC) on board Mars Global Surveyor (MGS).
MOC was also the first instrument to notice that areas where araneiforms have been found show substantial activity during local spring \citep{Malin_Edgett_2007}.
Since this discovery, multiple instruments observed polar areas and reported repeatedly occurring dark and bright fans, seasonal cracks and non-linear temporal variations of the surface albedo in these areas \citep{Piqueux_2008, Hansen_2010, Pommerol_2011}.
Among these instruments is the HiRISE on MRO.
Observations made by HiRISE are the starting point of this investigation because HiRISE is capable of resolving araneiforms at up to to 30 cm spatial scale, and HiRISE images can be used to create digital elevation maps (DEM) with 1~m spatial and 30~cm vertical resolution.

Araneiforms and seasonal features associated with them are believed to be related to the spring sublimation of a seasonal \cotwo{} ice layer.
This layer forms during late fall and winter, with a thickness reaching 0.5 -- 1~m depending on the particular location on Mars \citep{Smith_2001}.
\cotwo{} ice created under martian polar conditions is likely partially in the form of slab ice, i.e. the polycrystalline ice form of \cotwo{} that has very particular properties: it is translucent in the visible wavelength range and rather opaque in the thermal infrared \citep{Hansen_1997}.
Based on these special properties of \cotwo{} slab ice, H. Kieffer proposed a model to explain the appearance of fans in spring and the creation and modification of araneiforms \citep{Kieffer_2007}.
In short it goes as follows: sunlight penetrates through the complete \cotwo{} ice layer and warms up the underlying substrate.
At some point at the beginning of spring the solar energy input becomes large enough to start sublimation of \cotwo{} from underneath the slab.
The subliming \cotwo{} increases the pressure underneath the ice slab and when the pressure reaches a particular critical value the slab ice breaks, releasing pressurized \cotwo{} through the created opening.
As the flux of gas moves underneath the ice towards the opening it rips off substrate material and transports it out of the sub-ice cavity creating a jet of \cotwo{} gas with entrained sand and dust.
The substrate material has a size distribution with the heavier particles being deposited in fans and the light ones being injected into the atmosphere.

The observed araneiform troughs are presumed to be eroded by repeating the seasonal cryo-jet eruptions over many years \citep{Kieffer_2007, Piqueux_2003}.
Kieffer's model is generally accepted by the scientific community and has given rise to many recent publications about jets and fans \citep{Pommerol_2011, Hansen_2010, Portyankina_2010, Pilorget_2011, Thomas_2011, Portyankina_2017}.

Recently HiRISE reported first observations of seasonal \cotwo{} jets actively modifying the polar substrate: new dendritic troughs appeared 3 to 4 martian years ago and since are expanding and developing new tributaries \citep{Portyankina_2017}.
These new dendritic troughs are found in the southern hemisphere.
They are similar in the scale to northern furrows -- small shallow branching channels -- observed on northern dunes \citep{Bourke_2013, McKeown_2017}.
The same seasonal phenomena (fans, cracks, spring albedo variations, etc.) are observed near furrows, araneiforms, and new dendritic troughs \citep{Hansen_2010}.
However, furrows are erased in summer and develop on the dunes repeatedly every spring.
They do not develop into branching troughs because the dune material is too mobile to sustain the shape for more than a season.
Observations of furrows nevertheless are important for understanding the process of erosion by the gas flow.

Seasonal processes in martian polar regions have been extensively studied \citep{Piqueux_2003, Hansen_2010, Thomas_2011, Pommerol_2011, Kieffer_2007, Hansen_2013, Bourke_2013}.
The modeling efforts so far have concentrated on describing the seasonally repeated phenomena, mainly jets and fans, while araneiforms were considered merely as topographical features that do not change.
Thus, the impact they have on the evolution of the terrain over geological time scales is under-investigated.
Araneiforms, furrows, and the newest dendritic troughs offer the connection between long-scale geological processes and short-scale seasonal activity that is widely observed today.

Several authors have noticed the similarity of araneiform patterns to dendrites, specifically to fractal river patterns \citep{Piqueux_2003, Hansen_2010, Pommerol_2011, Kieffer_2007}.
Both processes involve erosion of substrate by a flow of a moving agent: water in the case of rivers, and pressurized gas in the case of araneiforms.
Yet, river erosion is fundamentally different in that it is governed by gravity.
The process of creation of the araneiform terrains is specific to Mars and has no direct terrestrial analogs.
Araneiforms are created by gaseous flow erosion which does not follow the topographical gradient, but rather the gradient of gas pressure inside the chamber underneath the ice layer.
Araneiforms often exhibit central symmetry and in several locations their troughs widen uphill  \citep{Hansen_2010}.
From the observational point of view, the only dendrites with central symmetry observed on Earth are either completely outside geophysical domain (electrical discharges, phase changes in metal alloys, neurons, etc.) or observed on frozen lakes and are entirely due to phase change of ice to liquid and do not involve surface erosion.

It's worth to note here that formation of araneiform patterns was investigated under laboratory settings by \cite{deVillier_2012}.
The authors simulated a transient air over-pressure above and within the granular medium of spherical silicate glass beads using Hele-Shaw cell set-up.
Multiple repetitions of creating over-pressure resulted in brunching patterns were visually similar to the observed centrally-symmetric martian araneiforms.
The authors mention that for correct comparison of laboratory simulations with real martian landforms creation process "there are significant uncertainties in the experiments {...} and much larger uncertainties on Mars".

Within this study, we sought to establish a link between active seasonal processes currently observed in the Martian polar areas and their impact on the geological record of the underlying surface.
To achieve this, we implemented a mathematical model for the formation and modification of drainage networks of araneiform terrains under present day martian conditions and compared its results with the observed Martian araneiform shapes.
This paper is a proof of concept that the morphological descriptions used for terrestrial drainage networks descriptions (described in Section \ref{subsection:terrestrial_definitions}) and the model (described in Section \ref{DLA}) can be adequately applied to the specific case of martian araneiforms.
Below we provide several examples of  parameters that can be used for quantitative description of araneiform  morphology.
Future work is to expand the application of these parameters to a larger numbers of araneiforms and more diverse selection of regions of interest and araneiform morphologies. 
Here, we lay a foundation to the future work that will aim to establish quantitative links between the model and observational parameters.

\subsection{Dendritic networks on Earth}

Dendritic networks (or fractal networks) are widely occurring in nature.
Multiple examples can be found in geomorphology and terrestrial biosphere: plants and trees, river drainage basins, human and animals cardiovascular systems, crystals, lightnings  and many more  \citep{Horsefield_1976, Mandelbrot_1982, Bunde_1996, Turcotte_1998, Pelletier_2000, Ghosh_2006, Miall_2013}.
Because many researchers notice similarity between araneiforms and terrestrial rivers in morphology but also in other physical properties such as being topographical features carved by eroding agent, we concentrate below on methods used in terrestrial hydrology.  
 
Vast research is dedicated to the analysis of patterns of terrestrial rivers. 
In this project, we use statistical laws and quantitative geomorphological methods developed by \cite{Horton_1945, Strahler_1957, Shreve_1966, Smart_1968} and others.

The appearance of drainage networks on Earth depends on the stream slope, amount of water discharge per unit time, periodicity of rainfall, and properties of the soil that the drainage network is carved into.
There are a number of modifications needed to adapt such a description to the case of cold jets erosion on Mars.
Stream slope is related to gravity being a driving force of river erosion.
For araneiforms it corresponds to the pressure difference between the cavity below the ice and the ambient atmospheric pressure.
The shape of the terrestrial drainage network is highly dependent on the properties of the soil the network is carved in \cite{Strahler_1957, deVente_2005}.
It was shown that soft soils tend to create a lot of material suspended in the flow.
In places where the flow slows down, the suspended material gets deposited in the little deviations of the flow from straight line thus creating meandering channels.
In contrast, materials more resistant to erosion (granite for example) tend to develop straighter tributaries and more dendritic shapes of drainage networks, but over longer time spans.
This observation may be applicable to the Martian substrates as well.
We observe different meandering pattern in the araneiforms on Mars (see examples in section \ref{sec:morphology}).
Based on analogy, araneiforms with more winding channels are expected to have larger sediment transport through the channels and probably are more mature structures compared to those with undeviating channels. 
If correct, this means substrate properties such as degree of consolidation might explain the diversity of the araneiform shapes.
To prove or disprove this connection meandering of araneiform channels must be a topic of future more in-depth study.

The examples of araneiforms presented in Figures \ref{fig:complex_spider} - \ref{fig:large_combo} were intentionally selected to illustrate the diversity of araneiform shapes.
This is in no way a complete coverage of parameter space consisting of sizes and complexity of the araneiform structures, their branching ratios, depth of troughs, and so on.
The aim of this exercise was to provide range of observational examples for the diffusion limited aggregation (DLA) model runs.

\section{Araneiform morphology}\label{sec:morphology}

The geomorphological study of araneiforms has, to-date, been restricted to only qualitative descriptions.
To be able to model the formation process however, quantitative descriptors of araneiform patterns are needed for comparison to the computer modeled structures. 

\subsection{Some definitions}\label{subsection:terrestrial_definitions}

We adopt the definition for ''tributary'' used in hydrology studies and use the Horton-Strahler stream order scheme with necessary adaptations for Martian applications.

The original definition states that a tributary is a river or stream that flows into a larger river or a lake. 
It uses the terminology associated with water flow natural for hydrological studies on Earth.
For our purpose of studying araneiforms on Mars it may be at times be confusing because gravitationally influenced flows of liquids are not part of their formation process. 
However, the term "tributary" is also widely used in dried river basins to define a channel in which water would flow seasonally or was flowing before.
We will use the term in a similar way to describe the channel through which \cotwo{} gas flows in spring.

\subsection{Rules for defining tributary order}\label{sec:dendrite_orders}

Selected araneiforms were analyzed following the Strahler stream ordering scheme \citep{Horton_1945, Strahler_1957}.
According to it, the araneiform pattern is divided into a set of tributaries of different degrees.
A first degree tributary is the one that has no tributaries, two first degree tributaries merging into one form second degree tributary, and so on.
The surface density of tributaries of different degrees, their absolute number and ratios are descriptive parameters of the drainage network that can be used for the comparison with the model results.

\begin{figure}
\includegraphics[width=0.45\columnwidth]{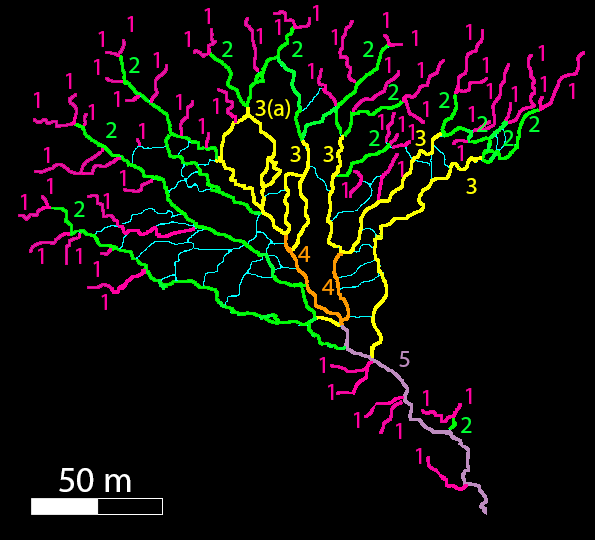}
\includegraphics[width=0.45\columnwidth]{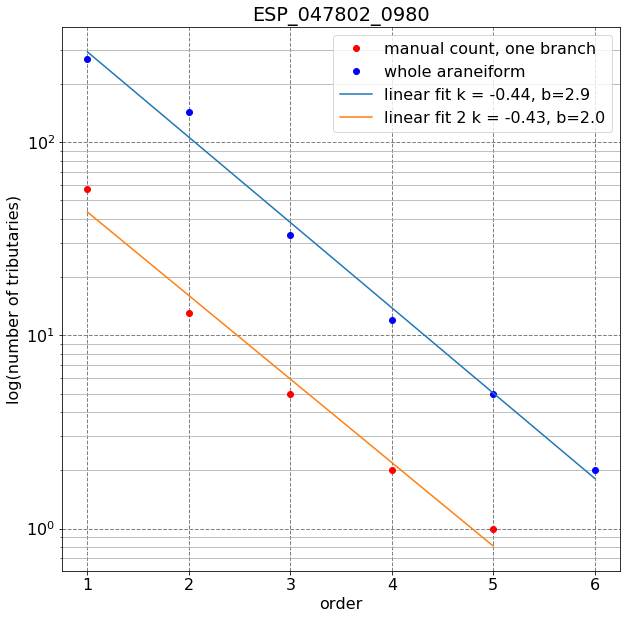}
\caption{\textbf{\label{fig:one_branch}} Left: Zoom on one branch of araneiform in Fig. \ref{fig:complex_spider} shows an example of morphological analysis done in this work. 
The orders of tributaries are marked in according colors.
The maximum order tributary is 5.    
Right: regression of number of tributaries on tributary order. According to Horton's law \citep{Horton_1945} number of tributaries tends to follow inverse geometric series. 
}
\end{figure}

Fig. \ref{fig:one_branch} illustrates the use of this scheme on a single branch of an araneiform with the left panel showing tributaries traced from original image used for this analysis.

We determined orders of tributaries (or Strahler number) for a range of araneiform patterns selected to showcase the diversity of observed araneiforms.
Order of a tributary of a mathematical tree is a numerical measure of its branching complexity.
Rules to define tributary order in a dendritic system are as follows:

1) If a tributary has no tributaries flowing into it, its order is one;

2) For gravitationally-driven systems counting tributaries follows gravitational gradient; for quasi-circular systems -- radial lines from perimeter towards center; for other system if the gravitational gradient is unknown, the direction is from narrow to wide tributaries;

3) When two tributaries of the same order \textit{i} merge, the order of resulting tributary is \textit{i + 1};

4) When a tributary of order \textit{i} splits, the order of all resulting tributaries stays the same and equal to \textit{i};
Consequently, if any of these tributaries merge again, the resulting tributary order stays the same, equal to \textit{i};
This rule stems from drainage areas: only two tributaries of the same order that feed from different source regions increase a potential flow of erosive agent and thus create a higher order tributary if merged;

5) If tributary of lower order merges into tributary of higher order, their orders do not change;

The idea behind ordering tributaries in hierarchical manner is that they serve as transport channels for the erosive agent (water or gas) from outer parts of the structure towards the center where the agent escapes (in the case of rivers, to the ocean, and in the case of araneiforms, into the atmosphere). 
Thus for centrally-symmetric araneiforms we will be tracing tributaries from outside towards the center. 
For asymmetric shapes, we have to find other signatures of the increased flow, one of them might be width of the tributaries themselves.
However, as we will be showing below, this is not always straight forward: sometimes tributaries cross each other or connect orthogonally to the main flow direction.
In such cases we often refrain from assigning an order to a tributary.

Number of different order tributaries, surface density of tributaries of different orders, lengths of different orders of tributaries and their ratios are used as morphological parameters for describing the drainage networks. 

\subsection{Example 1: araneiform of order 6 from ESP\_047802\_0980}

Left panel on Fig. \ref{fig:one_branch} shows a zoom on one branch of the araneiform from Fig. \ref{fig:complex_spider}.
Tributaries marked light blue are orthogonal to assumed direction of gas movement.
This renders their order undetermined and does not influence ordering of the rest of tributaries.
We have omitted such tributaries in our analysis.

Fig. \ref{fig:one_branch} is a tributaries ordering sketch illustrates the use of several rules from section \ref{sec:dendrite_orders}.
For example, order-3 tributary marked 3(a) splits and then merges again. 
Following rule 4, its order kept 3 until it merges with another order-3 tributary from the right side.
There are multiple order-1 tributaries that merge into higher order tributaries.
Note that their orders do not change after those mergers.
In total, this branch has 
57  tributaries of order 1, 
13  tributaries of order 2, 
5  tributaries of order 3, 
2  tributaries of order 4, 
and 1 order 5 tributary.
We have plotted these number in the standart logarithmic plot used to illustrate Horton's law of stream numbers.
According to Horton's law \citep{Horton_1945} for terrestrial river basins numbers of tributaries of different orders in a given basin  tend to follow an inverse geometric series.
As shown in right panel of Fig. \ref{fig:one_branch} this law is also fulfilled for Martian araneiforms. 
We plot both manual counts on a single branch from above and automatic counts of tributaries performed on the whole araneiform shown in Fig. \ref{fig:complex_spider}.
The linear approximation for the regression 

\begin{center}$
\log_{10} (N_{\omega})= k\omega + b  $
\end{center}

provides a regression coefficient $k$ which is related to bifurcation ratio: $R_b = log^{-1}k$ linked to the geometry of the basin.
It never can be below 2, because it reflects the ratio of number of tributaries of order $i$ to that of order $i+1$ averaged over the basin .
In out example, $R_b = 2.69$ which means that on average each tributary has between 2 and 3 tributaries flowing into it.

\begin{figure}
\includegraphics[width=0.45\columnwidth]{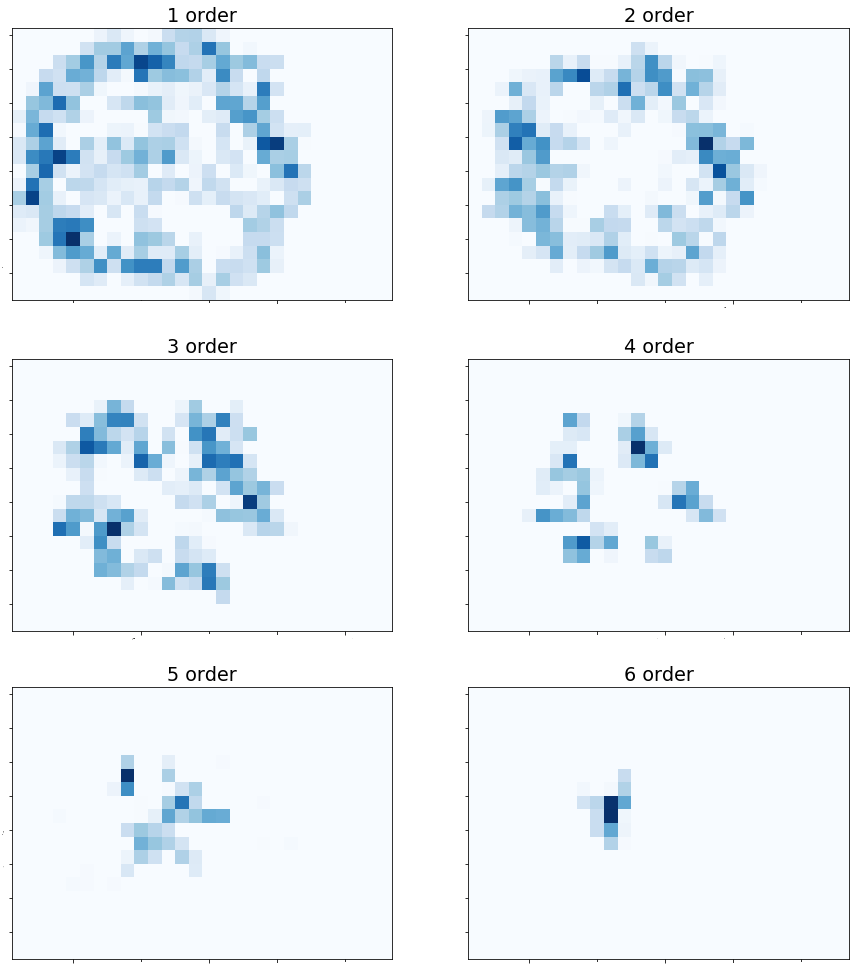}
\includegraphics[width=0.45\columnwidth]{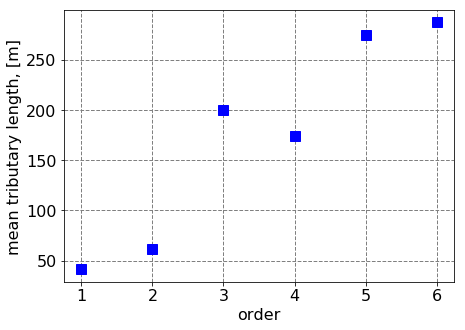}
\caption{\textbf{\label{fig:trib_den}}  Left: Normalized density maps of tributaries by order for the araneiform from HiRISE image ESP\_047802\_0980. Right: Mean tributary length plotted versus its order for the araneiform from HiRISE image ESP\_047802\_0980 shows an almost linear dependence. 
}
\end{figure}

Left panel of Fig. \ref{fig:trib_den} shows density maps for each order of the tributaries for the araneiform from Fig. \ref{fig:complex_spider}.
The densities are scaled at each frame to showcase the distribution of the tributaries in this structure.
First order tributaries are concentrated mostly near structure's perimeter, however some exist all over the araneiform, including close to its center.
All the following orders  tend to concentrate closer and closer to the center or in other words the order of tributaries on average increases radially towards the araneiform center.
Generally, the direction for increase of the tributary degree is from outer boundaries of the structure inwards.
Its hierarchical structure is centrally-symmetric which can be seen in density maps of different orders of tributaries (left panel of Fig. \ref{fig:trib_den})
In terrestrial river networks tributary degree increases following gravity, i.e. along the movement of the erosive agent.
For this araneiform the increase of tributary order towards center must indicate the movement of the erosive agent (\cotwo{} gas) is on average towards center of the structure.
This is however not always the case.
Figures \ref{fig:spiderESP-011630-0930} - \ref{fig:large_combo} include examples of araneiforms without central symmetry as well as large structures that developed secondary centers radially away from geometrical center of the structure.

The araneiform from Fig. \ref{fig:complex_spider} is a reasonably complex structure, its  maximum tributary order is 6. 
For comparison, Amazon river maximum order is 12, while 80\% of the streams on Earth are first to third order \citep{Downing_2012}.
The lengths of tributaries increase with tributary order for terrestrial basins.
We have plotted mean tributary length vs its order for the araneiform in Fig. \ref{fig:complex_spider}.
Generally, the trend is as expected: larger order tributaries are on average longer.  

All the above-mentioned analysis can only be done if an araneiform has statistically meaningful number of tributaries. 
Araneiforms come in a variety of forms: some of them have strongly developed branching of long troughs, some show only one or two troughs merging together, and others even show only one short arm connected to a pronounced center (many examples and some discussion of morphology can be found in  \cite{Hansen_2010}).
We have selected several examples to illustrate the variety of araneiforms from the side of tributary analysis.

\subsection{Example 2: large araneiform from ESP\_031793\_0980 }

\begin{figure}[h!]
\begin{center}
\includegraphics[width=0.85\columnwidth]{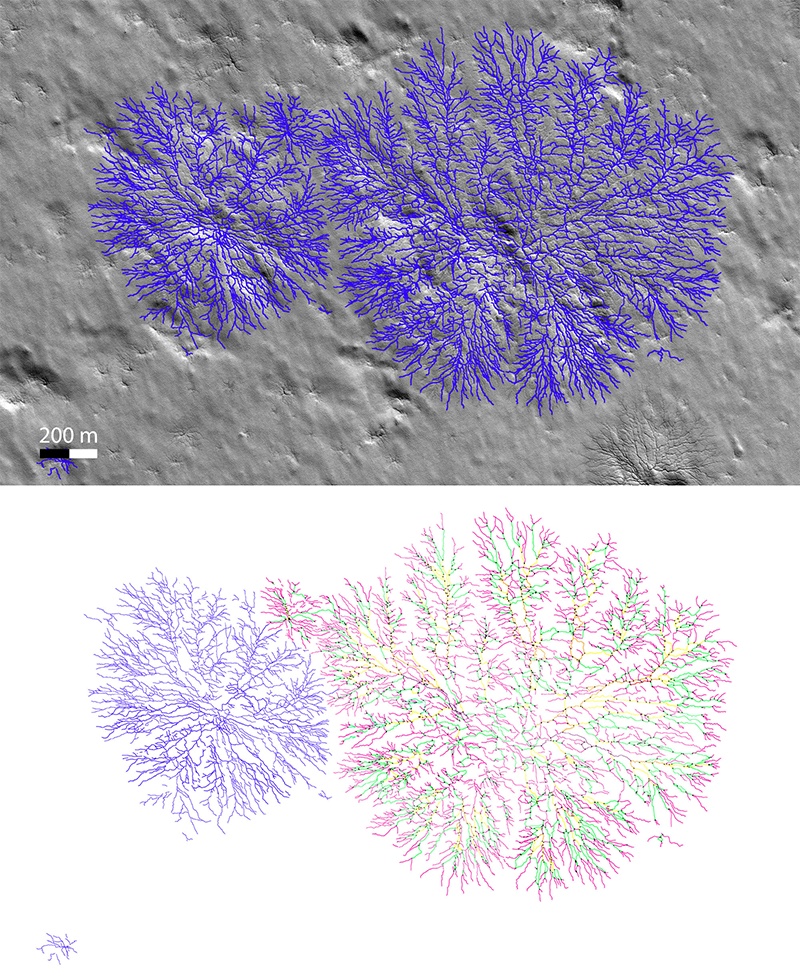}
\caption{\textbf{\label{fig:ESP0317930980}} Top: Sub-frame of HiRISE image ESP\_031793\_0980 with outlines of a large complex araneiform structure. Bottom: tributary order analysis of the largest araneiform structure in the frame. The tributary order is marked with color: 1st order is outlines in magenta, 2nd -- in green, 3d -- in yellow, 4th -- in orange, 5th -- in light purple, 6th -- purple. One can see the ''canopy shyness'' effect between the largest araneiform and the smaller one next to it.
}
\end{center}
\end{figure}

Image ESP\_031793\_0980 is taken at location informally dubbed Starburst at latitude -81.8\degree{}, east longitude 76.1\degree{}.
The surface here has multiple irregular pits overlayed by multiple very well developed large araneiforms.
Figure \ref{fig:ESP0317930980} shows several large and well developed araneiforms of many examples observed in this region.
We have drawn the outlines for 4 araneiforms in this frame: two large ones that dominate the central part of the image, a satellite araneiform that formed between the two, and a small one in the lower left corner that is not attached to the larger structures. 

\begin{figure}[h!]
\begin{center}
\includegraphics[width=0.52\columnwidth]{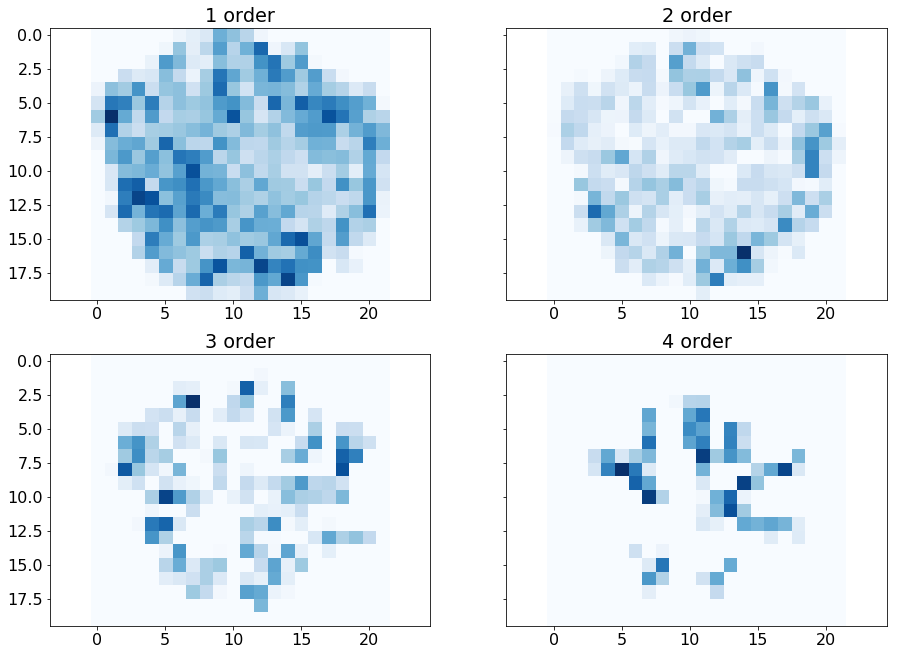}
\includegraphics[width=0.42\columnwidth]{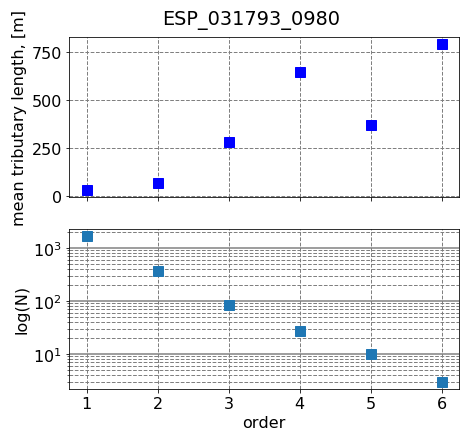}
\caption{\textbf{\label{fig:ESP0317930980_plots}} Left: Normalized density maps by tributary order for the araneiform from ESP\_031793\_0980.  Right: morphological analysis plots. a) Mean tributary length vs its order; b) Number of tributaries per order follow Horton's law of stream numbers. 
}
\end{center}
\end{figure}

The largest araneiform was then marked with different colors, each color reflects an order of each tributary. 
This araneiform is approximately 2~km across, the largest of our examples. 
Black dots mark merging points of two or more tributaries - these are the points after which the degree of a tributary changes.
Magenta color corresponds to the first degree tributaries, green - to the second degree, yellow - third, orange - forth, light purple - fifth, purple - sixth.
There is no tributaries of higher than sixth degree in this structure.
The tributary distribution of the largest araneiform from this frame was analyzed and the results are shown in Fig. \ref{fig:ESP0317930980_plots}.

Fig. \ref{fig:ESP0317930980_plots} shows normalized densities of each order of tributaries.
Unlike our first example, the first order tributaries are spread almost evenly over the structure, including its center.
This is probably because this araneiform is larger, older and more eroded than the example araneiform from ESP\_047802\_0980.
This also indicates that the erosive gas movement over this structure in spring is not fully centralized, it probably develops secondary centers and multiple escape vents and paths towards them.
HiRISE images of well-developed araneiforms often showed fan deposits that emerge away from araneiforms' centers \citep{Hansen_2010}. 
That is why de-centralized under-ice gas flow was expected. 
However, there is no proven link between currently observed seasonal fan deposits and long-term modification of araneiforms topography. 
The morphological analysis adds additional independent argument to the statement of de-centralized gas flow solely from the perspective of pattern structure development.

Tributary orders closely follow Horton's law (Fig. \ref{fig:ESP0317930980}).
The bifurcation ratio of this araneiform $R_b= 3.48$ is the largest of our examples.

It is worth to notice that the two araneiform structures show a phenomena called in biology ''canopy shyness'': branches of one structure seem to avoid getting too close to the branches of the other one \citep{Rudnicki_2002}. 
Similar to the biological analog of tree canopies, the reason for this phenomena is mostly deficiency of limited resource. 
In the case of trees it's sunlight or space available for unrestricted movement of branches during high winds.
In the case of centrally-symmetric araneiforms it's most probably the lack of under-ice gas pressure in the boundaries between the large structures like in this example.
However, this does not seem to be an absolute rule for all the araneiforms, for example, araneiform developed into lace terrain does not show this phenomena \citep{Hansen_2010}.

\subsection{Example 3: araneiform with central depression from ESP\_011776\_0930}

\begin{figure}
\includegraphics[width=0.35\columnwidth]{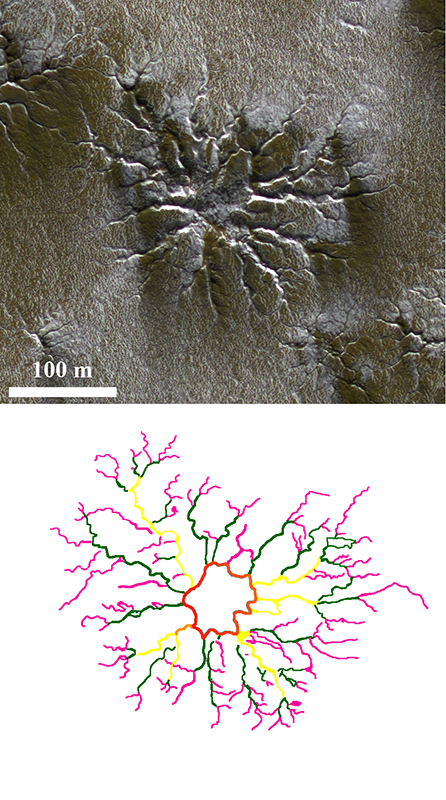}
\includegraphics[width=0.5\columnwidth]{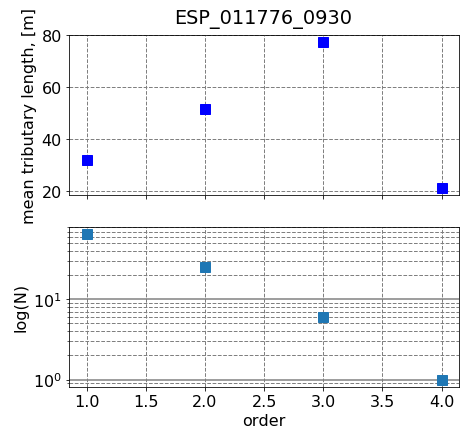}
\caption{\textbf{\label{fig:spider_morphology1ESP_011776}} Left: An example of morphological analysis of one araneiform from image \nolinkurl{ESP_011776_0930}. 1st order tributaries are outlined in magenta, 2nd -- in green, 3d -- in yellow, orange line shows the extend of central depression.  Right: morphological analysis plots: a) Mean tributary length vs its order; b) Number of tributaries per order follow Horton's law of stream numbers. 
}
\end{figure}

Image ESP\_011776\_0930 is taken at location  latitude -87.0\degree{}, longitude 127.3\degree{}.
Fig. \ref{fig:spider_morphology1ESP_011776} shows another example of what is considered to be a classical centrally-symmetrical araneiform.
The scale of this araneiform is smaller than our first example, it is about 250~m across, and over-all the system is less complex.
Maximum tributary order here is 4, and it has 
65 tributaries of order 1, 
25 tributaries of order 2, 
6 tributaries of order 3, 
and only 1 tributary of order 4. 

\begin{figure}
\includegraphics[width=0.65\columnwidth]{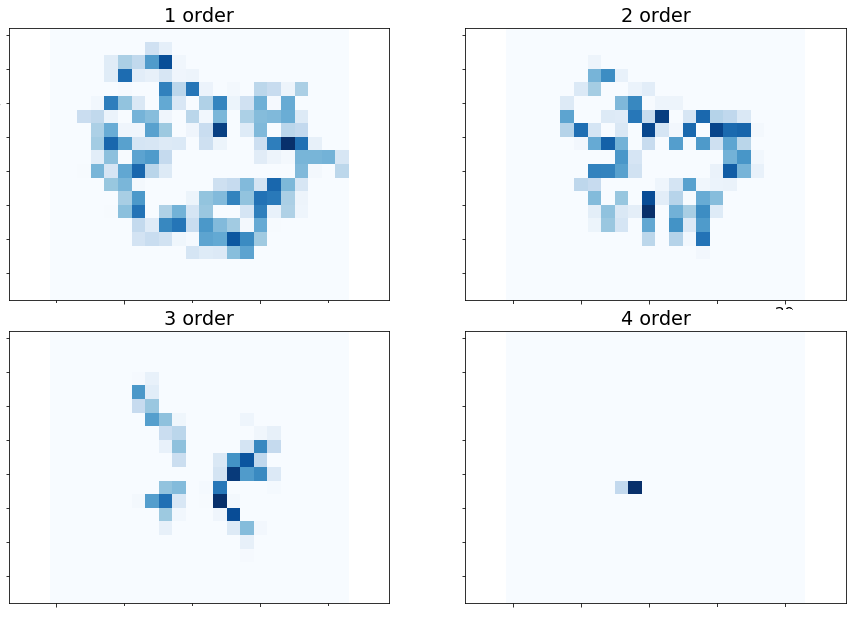}
\caption{\textbf{\label{fig:ESP_011776_den}} Normalized density maps of tributaries of different orders from image \nolinkurl{ESP_011776_0930}. 
}
\end{figure}

This araneiform  has large central depression where all tributaries terminate.
It is outlined in panel b) together with tributaries.
In proportion to the general araneiform size, the mean length of the tributaries here are smaller than in the first example but they tend to be wider, particularly of order 2 and larger. 
The bifurcation ratio is 3.2 calculated from plot e) (Fig. \ref{fig:spider_morphology1ESP_011776} ) -- slightly larger than in the first example.
The central symmetry is less obvious from visual appearance, mostly because of the presence of central depression.
The tributaries density maps (Fig. \ref{fig:ESP_011776_den}) show that the second order tributaries are more narrowly concentrated around the central depression than first-order tributaries which is consistent with central symmetry.
Third order however shows no particular pattern and forth order is not particularly useful because there is only one tributary.

\subsection{Example 4-10: some special cases}

All araneiforms shown below exhibit no central symmetry.
We have selected these examples to illustrate the diversity of araneiform patterns observed on Mars.

\begin{figure}[h!]
\includegraphics[width=0.35\columnwidth]{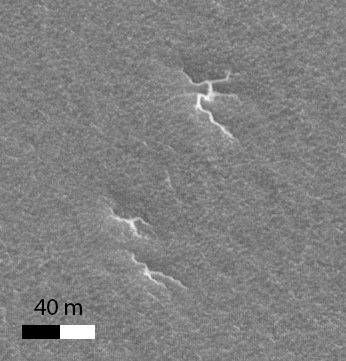}
\includegraphics[width=0.35\columnwidth]{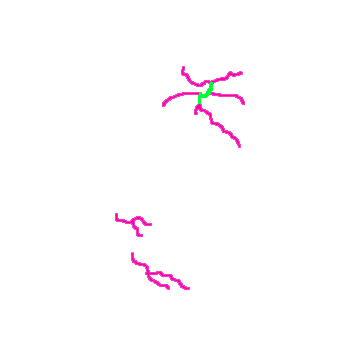}
\caption{\textbf{\label{fig:spiderESP-011630-0930}} Left: Sub-frame of HiRISE image ESP\_011630\_0930 showing araneiforms and their troughs outlined in blue. Right: Araneiforms' outlines without the HiRISE image background. Magenta shows first order tributaries, green shows second order tributaries. The maximum degree of tributaries here is 2. These araneiforms are likely in yearly stages of their development, i.e. with young apparent age. 
}
\end{figure}

Fig.~\ref{fig:spiderESP-011630-0930} shows an example of very small araneiforms that however have characteristic meandering tributaries.
The structures are small, not larger than 40~m across and consist from 3-5 tributaries.
The maximum tributary order here is 2. 
This is a good example of araneiforms with young apparent age.
It is up to now unclear why some areas have well developed araneiforms with old apparent age and others -- only small apparently younger structures, like this particular area (lat = -87.2\degree{}, lon = 167.9\degree{}).
This can either be indicative of less erosive substrate or that this surface was covered and unavailable for cold jet erosion.

\begin{figure}[h!]
\includegraphics[width=0.45\columnwidth]{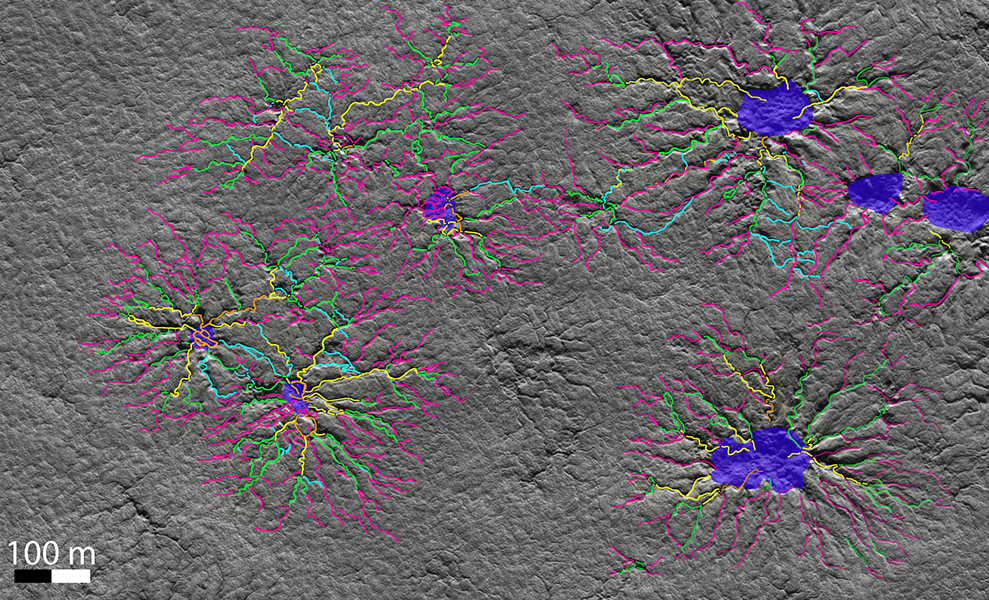}
\includegraphics[width=0.45\columnwidth]{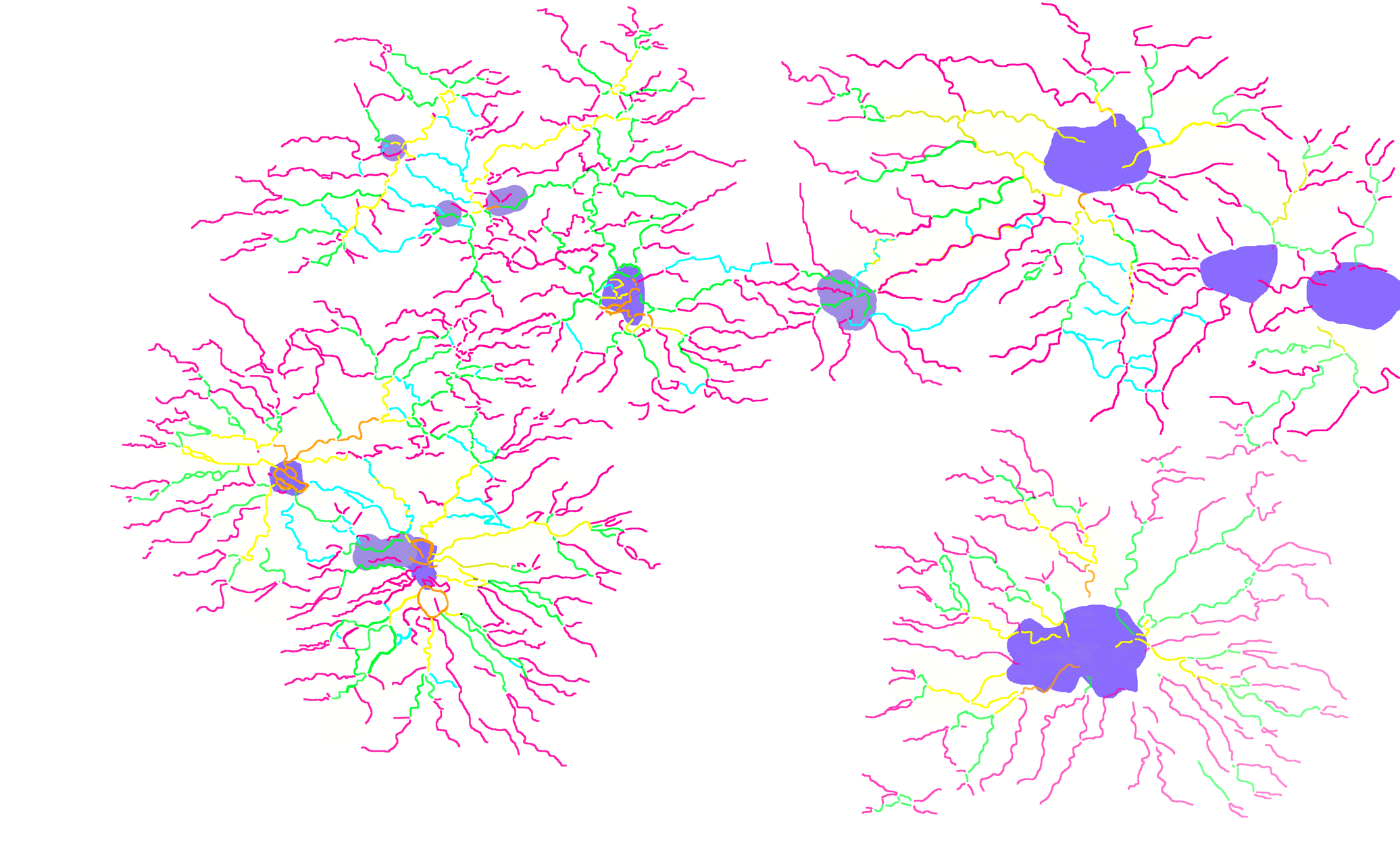}
\caption{\textbf{\label{fig:spiderESP-014282-0930}} Left: Subframe of HiRISE image ESP\_014282\_0930 showing araneiforms with central depression and their troughs outlined. This system is of order 4. As before, the tributary order is marked with color: 1st order is outlines in magenta, 2nd -- in green, 3d -- in yellow, 4th -- in orange. Light blue are tributaries of undefined order. Dark blue areas outline central depressions. Right: Araneiforms' outlines without the HiRISE image background. 
}
\end{figure}

Araneiforms on Fig.~\ref{fig:spiderESP-014282-0930} have very large central depressions and loosely ordered tributaries of maximum order 3 (outlines in yellow).
Araneiforms are up to 600~m across.
They overlap each other at places and have interconnected tributaries, i.e. not independent structures but rather form a common connected system. 
Large percentage of the tributaries of this structure are hard to assign to an order -- many are crossing each other or connecting the tributaries of the same order, meaning they did not participate in the transport of gas to one direction in a consistent manner.
The surface generally is uneven and eroded.
It is expected that the layer of seasonal ice draped over bumpy surface is prone to higher number of weak points.
These will serve as outgassing vents that randomly change their locations every spring and thus, if at all, create chaotically oriented tributaries.
The central depressions however provide more consistent locations for the vents and slightly focus erosive gas flow.
They will promote creation of tributaries running towards central depressions.

\begin{figure}[h!]
\includegraphics[width=0.45\columnwidth]{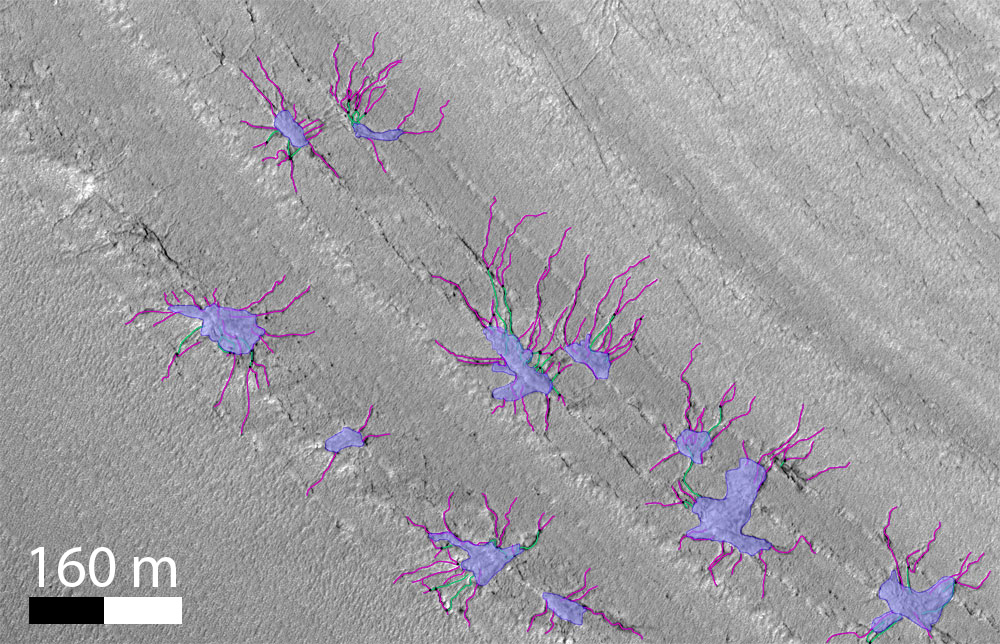}
\includegraphics[width=0.45\columnwidth]{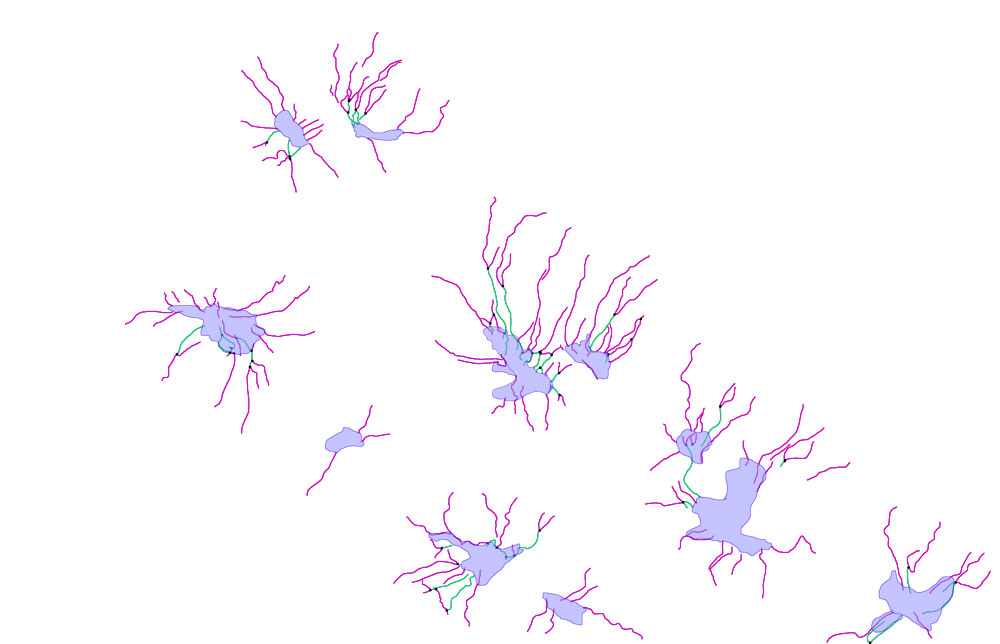}
\caption{\textbf{\label{fig:spiderPSP-005102-1030}} Left: Subframe of HiRISE image PSP\_005102\_1030. Araneiforms formed inside the SPLD. Maximum tributary order in this system is 2. 1st order tributaries marked in magenta, 2nd -- in green, central depressions -- in dark blue. Right: Araneiforms' outlines without the HiRISE image background. 
}
\end{figure}

Fig.~\ref{fig:spiderPSP-005102-1030} shows an example of araneiforms formed inside the layers of South polar layer deposits (SPLD).
Here araneiforms are less than 200~m across.
They form in two parallel SPLD layers at semi-equal distances and have pronounced irregular central depression.
The tributaries of maximum second order extend to both sides of the SPLD, i.e. both uphill and downhill.
Low order of araneiforms means young apparent age.
The rate of erosion of the outcrops strongly varies over the polar regions and  between southern and northern hemispheres, but generally, polar layer deposit outcrops, like the one imaged here, are considered to be examples of relatively young surfaces, i.e. 30--100 Ma \citep{Koutnik_2002, Carr_2010}.
Araneiforms formed on SPLD are obviously younger than SPLD themselves, which means the real, in addition to the apparent, age of these features is young.

\begin{figure}[h!]
\includegraphics[width=0.65\columnwidth]{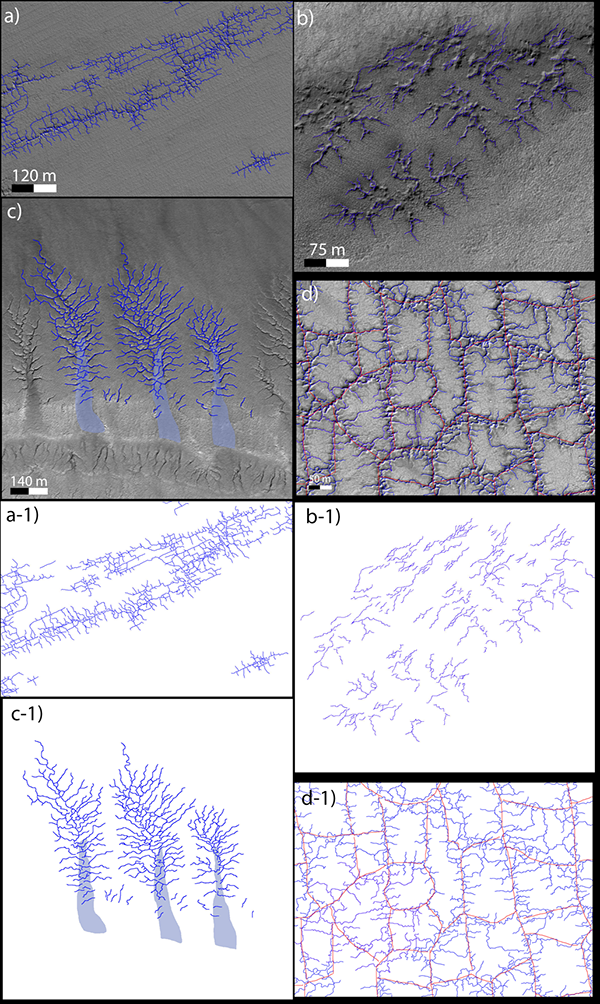}
\caption{\textbf{\label{fig:large_combo}} Examples of asymmetric araneiform pattern. a) Sub-frame of HiRISE image PSP\_005803\_0945 with outlined araneiforms developed over the polygonal crack terrain. b) Sub-frame of HiRISE image ESP\_040529\_0975 shows irregularly-shaped araneiforms with young apparent age. c) Sub-frame of HiRISE image PSP\_005627\_1015 shows very asymmetric araneiforms developed on foothills of polar layered deposits (PLD). d) Sub-frame of HiRISE image  ESP\_013824\_0950 showing polygonal araneiforms. The original polygonal terrain is outlines in red, and araneiform tributaries -- in blue. 
}
\end{figure}

And finally, Fig.~\ref{fig:large_combo} shows a set of 4 different examples of araneiforms.
Panels a and d show araneiforms with primitive tributary development.
These two examples is where araneiforms form over preexisting topography: either cracks in the top surface or polygonal terrain.
The tributaries run mostly orthogonal to the existing depressions (this is easier to see in the outlines of tributaries shown in panels a-1 and d-1 accordingly).
This means that the gas was escaping along those depressions during the formation of these systems.

Panel b shows a peculiar case where mostly separated channels are formed in a confined area, probably a former dune bed.
There is hardly couple of tributaries of order 2.
While some of the channels seem to group, they do not form a cohesive system of gas escape paths.
This might represent a frozen stage in development of a larger araneiform system.

Panel c shows another example of araneiforms formed near SPLD. 
The araneiforms are asymmetric along one axis which makes them look like tree-like structures.
The tributaries run towards an elongated depression that terminates at the beginning of higher layer. 
Neighbor araneiforms are evenly spaced.
 
\newpage

\newpage

\subsection{DLA application}\label{DLA}

Our premise for this work is a correspondence of the observational differences in araneiform patterns to different stages in their development: simpler looking and mostly smaller araneiforms are in the early stage of their development while the more complex ones are older.
We refer to this observed quality of different araneiforms as an ''apparent age''.
The correspondence between shape and apparent age is hypothesized and has not been independently verified simply because there is no means to estimate ages of small structures like araneiforms.
We introduce this term because it is hard to distinguish if a structure appears morphologically old because of its real age or because the conditions of its local environment are such that they promote fast tributary growth. 
With DLA modeling we aim to investigate if the distinction is possible.

There are several parameters that can influence the apparent age of an araneiform.
Probably the most influential parameter is substrate properties (such as material strength, compaction and cementation degree, content of the water ice, etc.) that directly determines how easy it is to erode the substrate.

Another important parameter is the erosion force of the sub-ice gas flow, which relates to the overall energy content of the jet eruption.
This energy in turn is controlled twice by the ice layer properties: once, for the amount of transmitted light, and secondly, because higher ice strength can store \cotwo{} gas at a higher pressure.

There is one parameter that influences the real age in contrast to the apparent age of araneiforms: if the retreat of the permanent polar cap was not symmetric then some areas were still covered at times when other areas already experienced the erosive forces of seasonal cap sublimation.
Because of this, differences in apparent age might have been caused by different exposure to the erosive forces.

It is hard to distinguish if a structure appears old because of its real age is old or ''apparent age'' is influenced by any or all of the listed parameters or  have from the apparent or real age of an araneiform.

\subsubsection{DLA model}

We have implemented a diffusion limited aggregation (DLA) model to describe the araneiform structures to investigate the relationship between the physical processes that are modifying araneiforms and their present appearance.
In this paper we present the model description and first applications to Martian araneiforms and furrows, and discuss how mathematical modeling of morphological structures helps to understand the complex system development.

The DLA model is a strictly mathematical model for pattern formation and was first introduced as a model for irreversible colloidal aggregation \citep{Witten_1983}.
It was later realized to be applicable to other naturally appearing patterns (like electrical discharge paths, mineral inclusions in rocks, spread of bacteria colonies, etc.) with the main common physical feature being that the transport of material is dominated by diffusion and not convection.
Most importantly, the parameters of the DLA model that influence the appearance of the resulting structure were shown to be related to the physical processes (erosion in our case) that create the structures observed.

A short description of the DLA model is as follows: consider a domain made of a 2-dimensional grid of points, one particle to serve as a non-moving seed for a future cluster and another particle to perform a random walk 
To allow for sufficient ''randomness'' the random walker should start in a significant distance from the clustering seed.
At each time step the direction of movement of the random walker is given by Monte Carlo test controlled by probabilities at each of six surrounding grids. 
At the end of its walk the walker might either have left the model domain or irreversibly adhered to the clustering seed.
If it adheres to the cluster, this growth of cluster represents the growth of the araneiform pattern and each adherence represents an incremental erosion event.
Repeating these random walks many times will make the cluster grow in size.
The clusters that are generated by such a model are both highly branched and fractal.

It is intriguing that the model can reproduce the geometrical pattern of naturally occurring systems but its relevance to describe those systems had to be proven.
It was shown that the mathematical description of DLA is remarkably similar to the one of Hele–Shaw fluid flow \citep{Witten_1983} that describes the spreading of water in an aqua-phobic liquid.

In a Hele-Shaw flow the pressure field in the fluid satisfies the Laplace equation.
For DLA modeling, the probability density of the random walker satisfies the Laplace equation.
In a Hele-Shaw flow, the boundary between the water and the aqua-phobic liquid is a constant-pressure boundary, and the velocity of the interface between the two liquids is proportional to the gradient of the pressure.
If water part inside Hele-Shaw cell is considered to be a cluster (the term used for DLA modeling), the velocity of the interface between the two liquids is now an analog for the probability of the "water cluster" growth. 
For DLA modeling,  the surface of the cluster is a constant probability density surface.
The probability of cluster growth on this surface is given by the gradient of probability density.
Thus the gradient of probability density can be compared to the pressure gradient in a Hele-Shaw flow.
Analog of pressure gradient to probability density effectively makes DLA a stochastic version of Hele-Shaw flow.

Based on the success of connecting DLA modeling to naturally occurring patterns \citet{FractalRivers} described the evolution of river basins using DLA.\@
Following in their steps, we apply DLA to araneiforms, based on the similarity of araneiform patterns to river basins.

\subsection{Basic implementation}

We have developed a DLA model for the creation of a 2D dendritic pattern similar to the appearance of araneiforms.
The algorithm is as described above.
In our implementation the integrated movements of all random walks are equivalent to the \cotwo{} gas flux underneath the seasonal ice layer.
The probability field that governs the movement of random walker is then equivalent to the gas pressure distribution underneath the ice.
The arrival of a random walker at and its attachment to the cluster is equivalent to an erosion event that extends a tributary either in width or length.
In the current implementation, the erosive event is always of the size of grid cell. 
This means that the resulting cluster can be freely scaled to the physical size of a particular araneiform and thus that the absolute scale bears no physical meaning and for now will be avoided in the plots.
We work on extending the current model to allow for variation of the erosive event radius, and thus introducing physical dimension into the ratios between the model grid and resulting cluster.
This is however beyond the scope of this paper.

Due to the stochastic-probabilistic nature of the DLA model we reduced the number of physical parameters that govern the morphological structure of a dendrite to the probability field that controls the random walker.
This reduction can be achieved by introducing dependencies of the probability field on time, physical space (instead of grid space) and physical properties of the surface and the \cotwo{} gas flow erosion.
We'd like to emphasize here that this is not for a single value for probability but rather a complex function of multiple physical parameters.

In the current implementation of DLA, we only consider cases with one original vent.
On Mars however we observed that multiple fans and thus eruptions of \cotwo{} may happen over a single araneiform.
This might look like a contradiction, however, at the moment when the first random walker attaches to the cluster, the vent definition becomes irrelevant. 
The outgassing may be considered to either continue through the original vent as if the gas would travel through the cluster towards it, or randomly outgassing from any of the cluster cells.
This flexibility is an outcome of the fact that for now we only consider a 2D structure, i.e. our cluster branches do not have depth.
In our future work we plan to consider the process of deepening of cluster branches which means we will have to follow the escaping gas along its way towards the vent.
This is a necessary step for 3D DLA modeling.
However, for the current work we will restrict ourselves to 2 dimensions. 

\subsection{DLA Example 1: centrally-symmetric probability field}

\begin{figure}
\includegraphics[width=1\columnwidth]{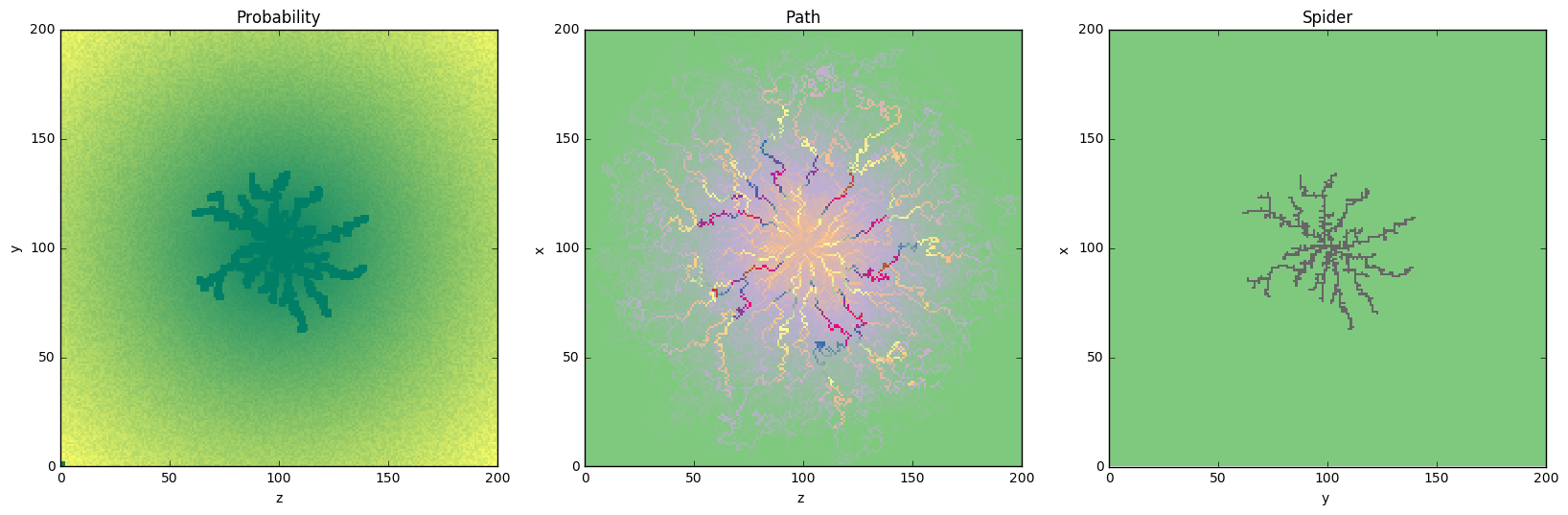}
\includegraphics[width=0.5\columnwidth]{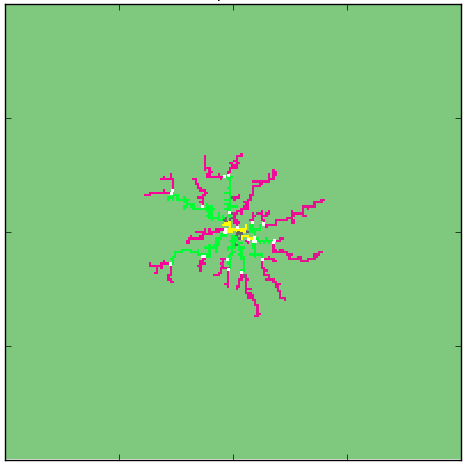}
\includegraphics[width=0.5\columnwidth]{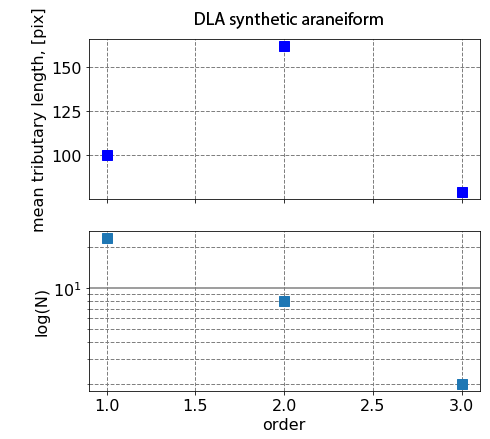}
\caption{\textbf{\label{fig:DLArun}} DLA run result for a centrally-symmetric probability field and morphological analysis of the acquired synthetic araneiform.
}
\end{figure}

The  implemented DLA algorithm works in 2 dimensions, i.e. it can create an ''image'' of a dendrite.
During the model run we call it cluster, and we'll refer to the resulting cluster as ''synthetic araneiform'' in contrast to the observed araneiform.
An example of a simple DLA model run is shown in Fig.~\ref{fig:DLArun}.
For this run we used a centrally symmetric probability field distributed over 200x200 pixels domain.
The starting seed of the cluster was placed in the center of the model domain.

The left panel of Fig.~\ref{fig:DLArun} shows governing probability field at the end of the model run.
The starting probability field followed radial dependency with an uniformed noise pattern overlaid on top of it.
The noise level can be changed depending on the desired degree of deviation of the corresponding gas flow from uniform.
The starting probability field f was modified during the run to slightly increase probability of the random walker towards newly developed araneiform branches.
This is visible as the cluster shape appears on the probability field plot with slightly thicker tributaries - this shows the range at which presence of an existing cluster influences the gas flow.

The middle panel is a combination of all random walkers paths and can be considered as analog of how the gas moved towards the central exit vent.
While the left panel can be put in analog to gas pressure, the middle panel is analog to gas flow.
We have chosen to plot a mid-sized cluster so that single paths of the random walkers that created it can be identified.
When the number of random walkers is increased and the cluster grows further, the path panel becomes saturated.

The right panel shows the resulting cluster or synthetic araneiform.
In this case the synthetic araneiform is visually similar to observed centrally-symmetric araneiforms like those from examples 1-3.
To support this statement, we performed morphologically analysis of the synthetic araneiform the same way we do for the observed araneiforms in section \ref{sec:morphology}.

This synthetic araneiform has maximum tributary order of 3, its tributary distribution follows Horton's law (Fig.~\ref{fig:DLArun}) and the bifurcation ratio was calculated to be $R_b = 3.39$, well inside the range of the observed centrally-symmetric araneiforms. 

In all examples of DLA model runs below we maintain the same plot arrangement.

Even the purely mathematical probability field provides plenty of room for variations.
We obtain different morphologies of cluster patterns by varying gradient of the probability field, its shape (central symmetry versus several minima and maxima), and the domain grid resolution.
In addition to the centrally-symmetric probability field, we have created two study cases to evaluate how these parameters influence the morphological shapes of modeled clusters.

\subsection{DLA Example 2: asymmetric araneiforms}

The obvious possibility to expand the parameter space for the DLA model is to deviate the governing probability field from centrally-symmetric.
The first study case of this type that we have created is to model araneiforms similar to those shown in Fig. \ref{fig:large_combo}.
These araneiforms are not centrally-symmetric and seem to either develop on some sort of pre-existing pattern (panels a and d) or have a strong asymmetric influence of some kind (panel c).
We have modified the governing probability field in our DLA run from circular to elliptic.
The gradient in this test was kept the same and the probability field shape was elliptical and aligned along the diagonal of our domain with the focuses in points (50, 50) and (100, 100) (panel a, Fig. \ref{fig:otherDLAruns1}), and the starting seed of the cluster was placed in position (100, 100).
The result of such DLA run is a cluster that is also elongated along the same axis with tributaries branching out to all sides (panel c, Fig. \ref{fig:otherDLAruns1}).
The cluster extends asymmetrically along the ellipse main axis: it shows more branching near the original seed and longer less branching tributaries towards the second focus of the probability ellipse. 
As one can see from panel b, random walkers start uniformly over the whole domain, but end up densely covering area around the forming cluster.
This is particularly obvious when compared to the centrally-symmetric case. 
As we noted above, integrated paths of random walkers in the case of araneiforms are representative of gas flux at the time of eruption.
The asymmetry in it indicates some kind of preferential pressure distribution repeated from year to year.
One possibility is that it is being created by the nearby topography change. 
For example, in the case from panel c of Fig. \ref{fig:large_combo} bottom part of the frame shows outcrop of polar layered deposits (PLD), i.e. araneiforms lay on the lowest layer of PLD and attach to the more inclined higher layer.
The asymmetry of the governing probability field together with the forming DLA cluster creates higher disturbances in the random walkers paths, meaning real gas flux near the forming araneiform will also experience higher disturbances.
Higher fluxes and larger disturbances lead to more efficient erosion.
This might explain why the araneiforms (like those from panel c of Fig. \ref{fig:large_combo}) can be observed near the edges of polar layered deposits despite these locations are considered to be one of most active places in Martian polar regions with scarps retreating and layers shed material downhill.

\subsection{DLA Example 3: polygonal araneiforms}

\begin{figure}
\includegraphics[width=1\columnwidth]{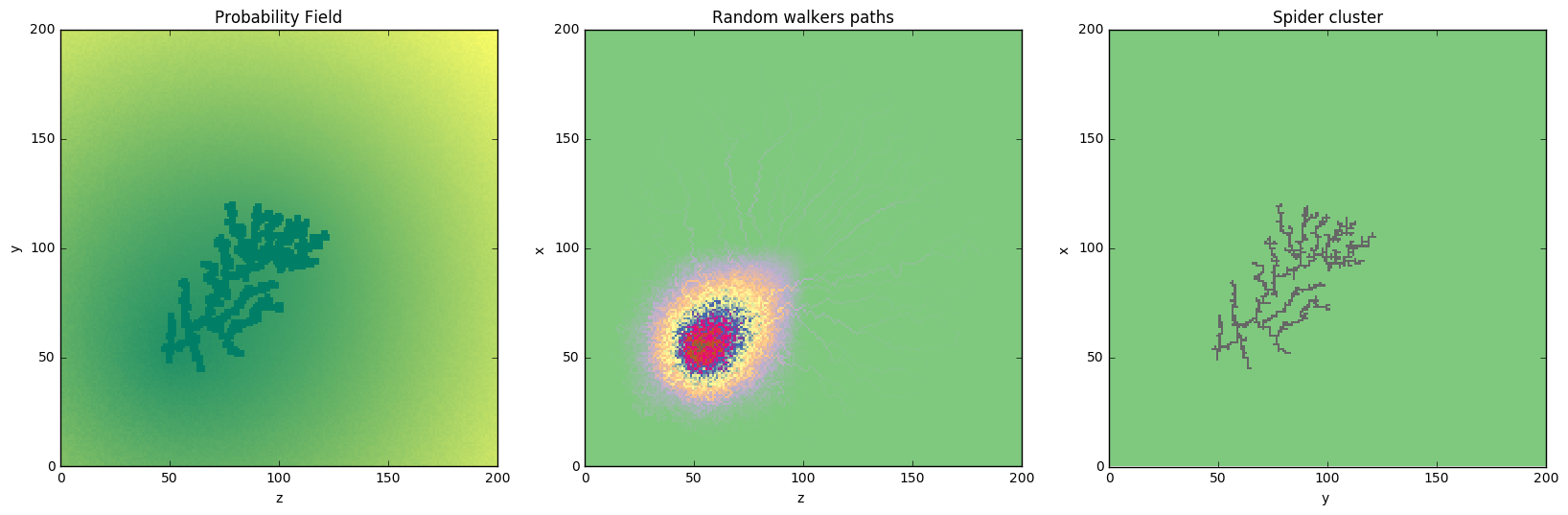}
\caption{\textbf{\label{fig:otherDLAruns1}} DLA run result for asymmetric probability field: a case of elliptical probability field with the main axis aligned on th ediagonal of the test domain. The resulting araneiform structure is elongated along the same diagonal.
}
\end{figure}

\begin{figure}
\includegraphics[width=1\columnwidth]{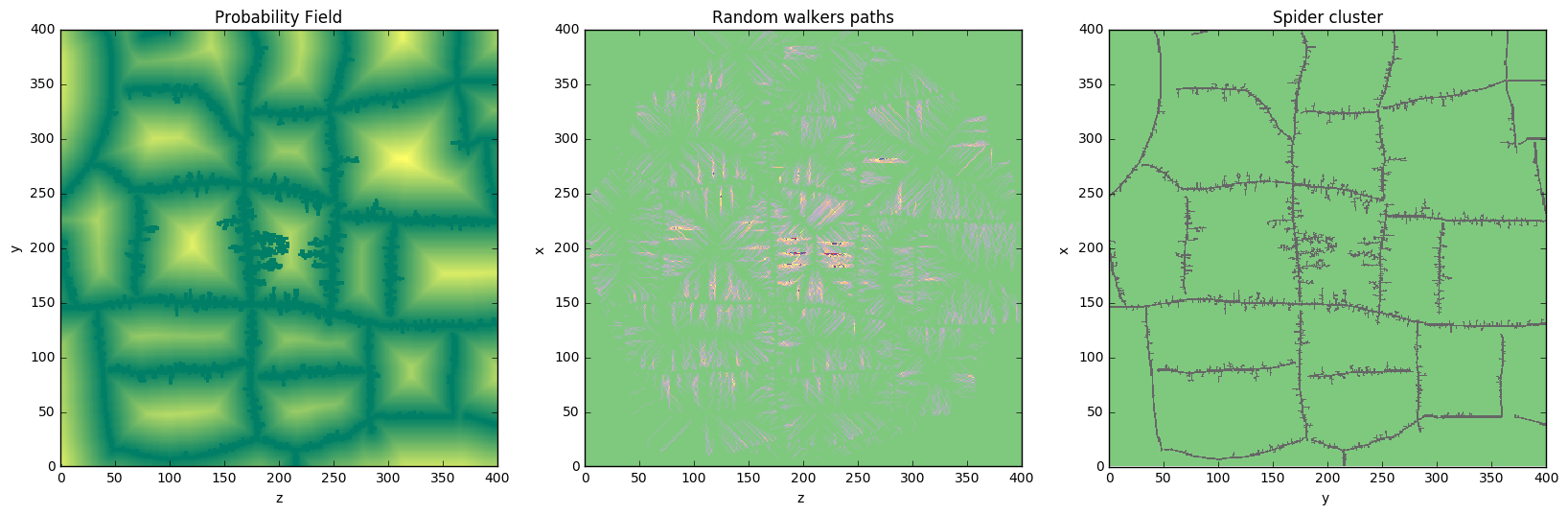}
\caption{\textbf{\label{fig:otherDLAruns2}} DLA run result for pre-existing polygonal terrain. Probability field in this case was  included polygons as location of increased probability. The resulting araneiform branches are orthogonal to polygonal lines, similar to the real araneiforms from panel $d-d'$ of Fig. \ref{fig:large_combo}.
}
\end{figure}

We have created our third test case based on observations of polygonal araneiforms (panel d from Fig. \ref{fig:otherDLAruns1}).
Polygonal terrains exist on Martian surface at virtually all latitudes.
Many of these terrains inside the polar regions have araneiform draping over them (Fig. \ref{fig:large_combo}, panels a and d), while most of those outside the polar regions do not.
We speculate that the araneiform development starts after the polygonal patterns appear. 
The analog to this disturbance in the substrate for the DLA model is the probability field having same pattern before the random walks.
We have traced an example of polygons from Martian surface and used this pattern as a starting DLA model probability field.
The size of domain in this case was increased to 400 by 400 to be able to incorporate several polygons of the pattern.
Fig. \ref{fig:otherDLAruns2} shows the resulting DLA run.
Only those random walkers that fall into the square will add to the tributaries in it, resulting in the high number of random walkers required to create the pattern shown in panel c of Fig. \ref{fig:otherDLAruns2}.
In this particular example we have used 5 times more randoms walkers compared to example 2. 
This is not only because of the domain increase, but also because density of the pattern is higher in the case of polygonal araneiforms.

On the final cluster panel (panel c) one can see that multiple tributaries extend from polygonal lines approximately orthogonally.
The longest tributaries develop close to the middle of the polygons, while short and not branching tributaries appear everywhere along the polygonal lines.
These properties are remarkably similar to the real polygonal araneiforms (Fig. \ref{fig:large_combo}, panels a and d). 

\section{Conclusions and future work}

We have for the first time applied qualitative morphological analysis to Martian araneiforms.
We showed that the methods used in terrestrial hydrology are applicable for the case of uniquely Martian terrains with some required modifications. 
The main modification in defining the orders of tributaries is that instead of following the gravitational gradient as is done in the case of rivers, we have to utilize symmetry of the pattern and trace tributaries either towards the structure's center or towards the direction of visibly thicker channels.
This is not always possible and many cases of araneiform patterns can not be analyzed this way, like those with young apparent age or those that developed around pre-existing surface structures.

Our morphological analysis of large developed araneiforms (with tributary orders larger than 4) shows that as dendrites they are structurally similar to terrestrial rivers.
They closely follow Horton's law of tributary orders and have bifurcation ratio that falls well inside the range of terrestrial rivers, which supports the hypothesis that both patterns are created by the erosive force of flowing agent on the underlying substrate.
\cite{Shreve_1966} has argued that deviation from Horton's law in terrestrial rivers are attributed to the irregularity of the substrate inside the river's basin and that the opposite is also true: the better a river's tributaries follow Horton's law, the more regular and smooth terrain the river is flowing through.
If we adopt the same logic for Martian araneiforms, we must conclude that all centrally-symmetric araneiforms developed in the areas with uniform substrate, i.e. substrate with uniform resistance to erosion at least in its top several meters that corresponds to the depths of araneiform troughs. 

\cite{Shreve_1966} concluded that ''populations of natural channel networks developed in the absence of geologic controls are topologically random {..} and therefore that, as proposed the law of stream numbers is indeed largely a consequence of random development of the topology of channel networks according to the laws of chance.
Thus, based on morphological similarity of araneiforms and rivers we conclude that the purely statistical models for pattern development, like diffusion limited aggregation (DLA) model, can be directly applied to study the formation of araneiforms.

We have implemented a two-dimensional DLA model that describes formation of different shapes of dendrites by mathematical probabilistic means.
We compared modeled dendrite shapes to the araneiform shapes observed in the Martian polar regions and evaluated their similarity using the morphological analysis of araneiforms.
We intentionally avoided using fractal dimensions of the observed araneiforms. 
Fractal dimension is a useful geometrical tool to describe a dendritic pattern but it cannot be solely used to evaluate quality of model fits mostly because different natural fractals can have the same fractal dimension down to third meaningful digit precision. 
It is possible that fractal dimension in connection with other morphological descriptors would provide a useful differentiating tool for the pattern description. 
This requires additional statistical investigations that are outside the scope of this work.

We have showed that it is possible to model different araneiform shapes with DLA model by modifying its governing probability field.
The analog to the governing probability field in the case of araneiforms is the pressure right before the under-ice movement starts.
The analog to the integrated random walkers field is the flux of under-ice gas flow. 
This is similar to how \cite{Masek_1993} interpreted random walkers for the case of terrestrial rivers.

The asymmetry of the governing probability field is required to create asymmetric araneiforms, which means that similar asymmetry in the gas pressure must be repeated from year to year for formation of many asymmetric araneiforms we have observed on Mars.
The pressure asymmetry may be because of the influence of nearby scarps of polar layer deposits or even simply change in inclination of the surface.
Spatial variations in surface inclination tend to provide weak spots in the ice layer readily creating vents early in the spring season.
This means, presence of inclination variations influences the gas pressure distribution underneath the ice every spring. 
Further studies of this are required to fully understand the influence of local topography on the development of araneiforms.

Our initial modeling indicates that the araneiforms near PLD scarps might be the fastest developing and also the youngest araneiforms.
This would fit to the idea that there should be no mature/well-developed araneiforms near PLD because it is very active areas with rock fall and increased sublimation of water ice from steep PLD slopes.

DLA model runs that included preexisting structure of polygonal terrains returned araneiform structures similar to those observed. 
This suggests that araneiform development may exploit preexisting polygonal cracks.

Our DLA model is in the early stage of development and multiple improvements to it are plausible. 
We plan to further develop our model and the first step is to introduce the scale of erosion events relative to the resolution of the model domain.
This will allow us to more realistically study geometry of tributaries, in particular their width to length ratios. 
We will use the examples of the currently ongoing erosion, like furrows and newly developed araneiforms, to scale the polar erosion in different regions and thus to study the geological history of the surface.

The other very important improvement to the DLA model is to expand it to three dimensional space.
This will allow us to better understand the erosion process and in the future to be able to estimate the real age of particular araneiforms.
This is an improvement to only operating with apparent ages that we have done until now. 
We will utilize the readily available digital elevation models produced by HiRISE that are able to resolve the largest araneiforms.

Over-all, this work is the first attempt on quantitatively analyzing the current Martian polar surface erosion and the formation of unique araneiform patterns.

\textit{Acknowledgments:} This project was supported  by the NASA grant nr. NNX15AH36G. The authors thank HiRISE team members for multiple useful discussions and continuous support.
The authors thank S. Piqueux and S. Diniega for their comments that helped to improve the manuscript.

\newpage
\bibliographystyle{elsarticle-harv} 
\bibliography{anyas_references}


\end{document}